# Advances in Spectral Classification

Sunetra Giridhar *

*Indian Institute of Astrophysics, Bangalore 560 034*

2018 August 11

**Abstract.**

In this article we give an overview of the developments in the field of spectral classification and its continued importance in the fields of stellar and galactic evolution. The extension of MK system to cool stars as well as refined classification of the hot stars using the new data obtained from modern ground based facilities and space missions has been described. A brief summary of automated methods of spectral classifications developed for the quick and objective classification of the ongoing and future surveys is presented. A new spectral class encoding system developed for exploitation of databases is described. Recent large scale spectral classification catalogues are listed.

## 1. Introduction

Although this review is aimed to introduce contemporary developments in this field, the subject of spectral classification and its impact on stellar and galactic astronomy cannot be introduced and appreciated without a brief mention of contributions made by pioneers in the field.

A. Secchi had observed and classified nearly 4000 stars (during 1863-70) firstly in two groups (based upon the strengths of hydrogen lines) and subsequently two more groups were added based upon the strengths of molecular bands. The credit for identifying carbon stars with spectral bands corresponding to laboratory spectrum of carbon also goes to Secchi. M.L. Rutherford on the other hand, not only developed refined accessories for acquiring stellar spectra, but also developed a classification scheme based upon stellar colours which later evolved into a powerful photometric system of classification. W. Huggins worked on the identification of stellar lines by comparing them with laboratory spark spectra. He was interested in stellar compositions and the conditions responsible for the observed spectra. G.B. Airy pursued a program to measure Doppler motions of the stars by recording the line positions for a large number of stars.

---

*e-mail:giridhar@iiap.res.in

## 1.1    Draper Catalogue of stellar Spectra

From 1886, Henry Draper Memorial Survey at Harvard carried out a systematic photographic spectroscopy of stars brighter than the 9th magnitude covering the full sky using telescopes at Harvard, USA and Arequipa, Peru under the leadership of E.C. Pickering. The spectral classes of A. Secchi were subdivided into more specific classes by W. Fleming by assigning spectral classes A-N depending (again) upon the strengths of hydrogen lines with A type stars exhibiting the strongest hydrogen lines. This system was found to be unsatisfactory since it resulted in a sequence where other lines and B-V colours showed irregular variations. This system was improved by A. Maury, A.J. Cannon and and E.C. Pickering into a natural sequence following the variations in other lines as well it was also a sequence according to the stellar colours. The resulting spectrum sequence OBAFGKM (notwithstanding the awkward arrangement of letters) is a stable temperature sequence and has been adopted. Cannon also exploited the larger database and improved quality of spectra of HD survey to subdivide the spectral types into decimal types like A0, A1, A2 ...A9. At the cool end, parallel branches of R, N, S stars are found with temperatures similar to those of M stars. While the M stars exhibit bands of TiO, S stars display those of ZrO. R and N on the other hand are carbon stars showing strong bands of carbon based molecules like CH, CN and $C_2$.

The Henry Draper (HD) Catalogue was published as Annals of Harvard college Observatory between 1918-1924. This monumental work provides rough positions, magnitudes, and spectral classifications of 225,300 stars. The extension to HD catalogue (HDE) was made under the leadership of H. Shapley providing additional spectral types to 46,850 for fainter stars reaching about 11th magnitude.

## 1.2    Two Dimensional MK System

In 1905, E.Hertzsprung noted a relation between the spectral width of certain lines with intrinsic luminosity of stars. Investigation of stars in various stellar systems by W.S. Adams, W.H.S. Monack, E.Hertzsprung and H.N. Russell led to a better understanding of relation between intrinsic luminosity of a star, its proper motion and parallax. It was reasoned that at a given apparent magnitude, the low proper motion stars would be at larger distances compared to those with high proper motions and hence would be of higher intrinsic luminosity. These investigators identified the lines which were sensitive to luminosities and ratios of lines which were sensitive to luminosity with those insensitive to luminosity were calibrated against its absolute magnitude (derived from parallax). This spectroscopic method of determining stellar parallax (Spectroscopic Parallax) had been used extensively to large samples of stars by W.S. Adams and his collaborators and became a powerful tool in the study of galactic structure.

In 1943, W. Morgan, P.C.Keenon and E. Kellman introduced MKK atlas comprising of 55 spectra. These photographic prints clearly demonstrated the luminosity effects and introduced luminosity as the second classification parameter. Morgan had observed the near constancy of the gravity along the main sequence in Temperature - Gravity diagram and luminosity class parameter was introduced to identify stars of different gravities and hence radii at nearly constant temperature.

The above mentioned system is also known as Yerkes Spectral Classification. Within the system, six luminosity classes are defined on the basis of standard stars over the observed luminosity range.

The Six classes are:

Ia: most luminous supergiants

Ib: less luminous supergiants

II: luminous giants

III: normal giants

IV: subgiants

V: main sequence stars

However, additional luminosity classes of 0 for hypergiants , VI subdwarfs and VII white dwarfs have been introduced in recent times.

The luminosity effects are not restricted to the narrowing of strong lines. The line strengths and ratios of line strengths of neutral and ionized species also show remarkable variations over spectral classes and luminosity types and have been used for defining the subclasses and luminosity types. The MKK atlas not only introduced refinement in the classification system, but also provided a large set of standard spectra.

Within a decade of its introduction, MK classification provided a mean of tracing local spiral structure in the Galaxy through the absolute magnitudes derived for B stars by Morgan, Whitford and Code (1953). W.W. Morgan together with H.L. Johnson not only introduced the broad band U,B,V system (Johnson and Morgan 1953), but also presented its calibration in terms of spectral types through observations of a large number of field stars as well as those in open clusters. The two colour diagrams and reddening correction procedure led to more accurate absolute magnitude and hence distances for luminous B stars leading to the discovery of Sagittarius-Carina arm and more detailed local spiral structure by Humphrey(1976)

Another important discovery was by N.G. Roman who found a group of weak-lined dwarfs and giants while classifying a sample of F,G,K stars. These weak-lined (metal-poor) also exhibited large velocities and were identified as population II objects. This finding provided basic observational data to develop theoretical models for the formation and evolution of our Galaxy.

### 1.3    MK Processes, Its Autonomy

The MK system has been built on spectral appearance of standard stars. The philosophy is based upon comparing the morphology of the spectrum to be classified with that of standard stars. This process is independent of other information such as colours and calibration of stellar parameters based upon theory. This autonomous nature has provided it with long term stability. Although it is known that spectral types are related to the temperatures and luminosity classes to gravities, the spectral classes are not assigned based upon these parameters but rather on the spectral appearance and their close match with the standard stars. Standard stars therefore play very important role in the classification procedures.

### 1.4    MK Standards

As we have mentioned earlier, the MK classification is made by comparing the unclassified star spectrum with those of MK standards. It is imperative that the standard stars are available for all spectral types and luminosity classes. The standard star list which was first provided with the MKK atlas has been expanded and refined by Keenan and McNeil (1976), Morgan, Abt & Tapscott(1978) and Keenan & McNeil (1989). However, a repository of standard stars using high quality (S/N ratio) spectra has been developed through dedicated efforts by R.F.Garrison, N.R. Walborn and R.O. Gray. A subset of standard stars which has survived scrutiny through better spectra with no changes in their status represent most stable (anchor points) of the MK system has been given in Garrison (1994). The introduction of new standards have not only improved the accuracy of classification procedure, but has also helped the extension of MK system to the L and T dwarfs. The standard stars are available now and are fairly well distributed over the sky. The establishment of fainter secondary standards more suitable for larger telescopes has been prime objective of the *Standard Star Working Group of the International Astro-* *nomical Union.* Several lists of standard stars dealing with different spectral types have been developed.

## 2. Revision and extension of MK system

The MK classification has been continuously upgraded incorporating the additional information brought in through larger telescopes, modern detectors and space missions. Even with the older accessories, the refinement to the MK system by introducing a third dimension such as metallicity was proposed by Keenan (1985). This work was followed up by Gray (1989) and Corbally(1987) by identifying spectral features suitable for two dimensional classification of metal-poor F,G stars and also providing a list of metal-poor standard stars. The refinement in spectral classification has also been made through the usage of the strengths of emission features. In the following subsections we will present the classification criteria used in the present times including the classifiers in ultraviolet (UV) and infrared (IR) spectral regions.

### 2.1 O stars

Being young and luminous objects, O stars are important tracers of sprial structure in our and external galaxies. The original MKK system of 1943 had standard stars defining only spectral range O9 to M2. The early attempts of classification of OB stars made use of lines strengths of He II and He I absorption lines. The re-examination of the spectra by Walborn and his collaborators led to the usage of better classification criteria such as Si IV, Si III and intermediate spectral types like O9.7, B0.2 were introduced. Stars as early as O3 were identified (Walborn 2002). The classification criteria were not restricted to absorption line strengths and their ratios but observed large variations in emission line strengths of N IV, N III, Si IV lines in OB spectral types were exploited. The spectral type (temperature) sequence is illustrated in Fig. 1 wherein large variations in spectral line strengths can be seen over spectral types O2 to O5.

The primary classification criteria for luminosity types had been ratios Si IV $\lambda4089$ / He I $\lambda4026$ or Si IV $\lambda4116$/ He I $\lambda4121$ for late O and early B and Si III $\lambda4552$/He I $\lambda4387$ to late B classes.

Most interesting development was realization that He II $\lambda4686$ and N III $\lambda\lambda$ 4634-4640-4642 features changed from absorption at dwarf luminosity class to emission for supergiants at the same spectral type. This luminosity effect has been illustrated in Fig. 2 for spectral type O6.5.

The lack of this understanding had resulted in O supergiant with emission features being considered peculiar O (Of) stars and were assigned labels ((f)), (f) and f for transitions from weak emission with strong absorption to pure strong emission feature. An on-line catalogue of classified O type stars and supplementary data from UV and IR is maintained by Maíz-Apellániz et al. (2004).

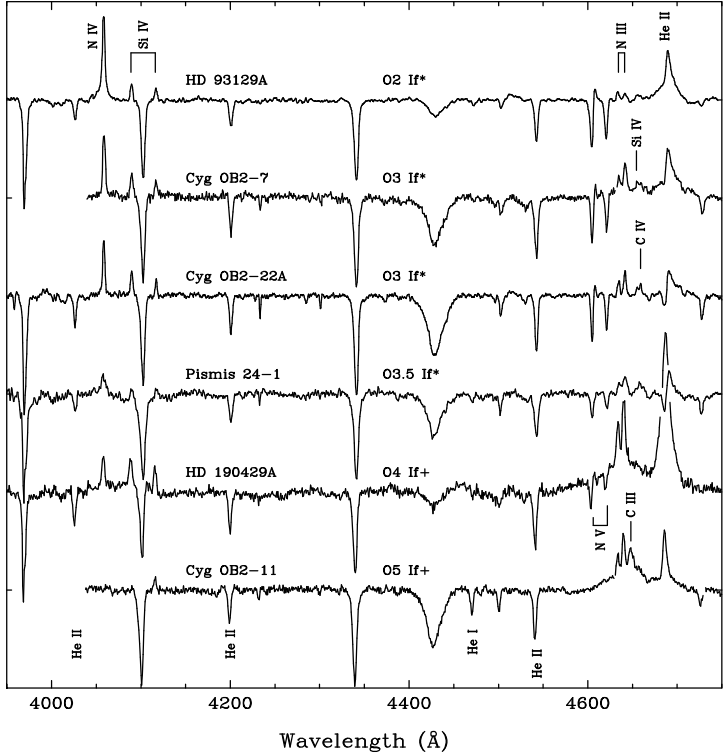

**Figure 1.** The temperature sequence for early O type supergiants. ( Figure courtesy I.D. Howarth and Editorial Complutense S.A. Reproduced from *UV Astronomy: Stars from Birth to Death* with permission from Editorial Complutense, 2007).

### 2.1.1  *UV Spectra of O stars*

Although it was known from the study of blue-visual photographic spectra that the O type luminosity sequence towards low gravity are also progression in stellar wind densities, the UV spectra of OB stars show much stronger signatures of stellar wind through the P-Cygni profiles of many lines e.g. N IV $\lambda\lambda$ 1239, 1243, Si IV $\lambda\lambda$ 1394, 1403, C IV $\lambda\lambda$ 1548, 1551 etc. It is interesting to note that although the UV spectral features of OB stars are dominated by wind features, they correlate very well with the spectral sequence built up using blue-visual spectra.

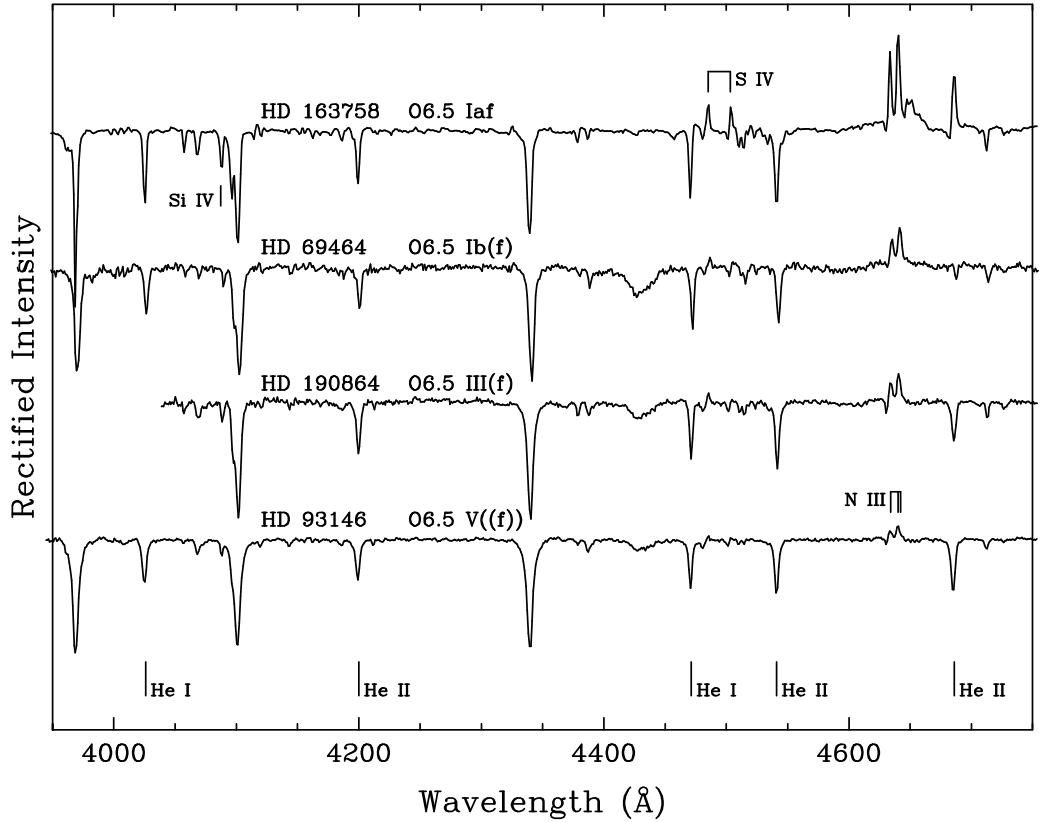

**Figure 2.** The transition from absorption to emission with increasing luminosity is illustrated for O6.5 stars. ( Figure courtesy I.D. Howarth and Editorial Complutense S.A. Reproduced from *UV Astronomy: Stars from Birth to Death* with permission from Editorial Complutense, 2007).

### 2.1.2 *IR Spectra of O stars*

The IR spectra also contains the features of He II, He I and C IV and they are used for classification. However, it spectra contains much fewer and weaker features than optical or UV regions. IR classification therefore requires resolution R > 1000 and S/N ratio of 70 to 100. The extinction at IR wavelength is much less than that at visual and UV. The IR classification is therefore very useful in the study of massive stars in young star forming regions, primary stars of X-ray binary and normal O stars hidden in the Galactic spiral arms as well as towards the Galactic center.

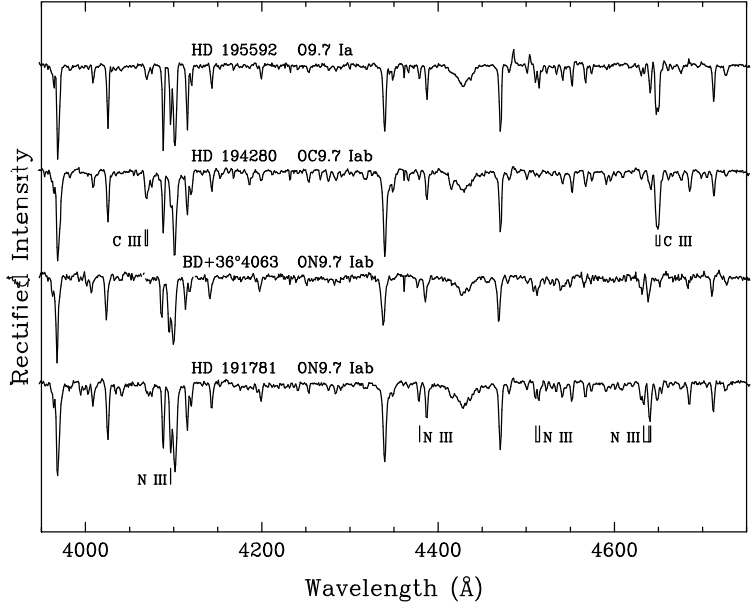

**Figure 3.** The C,N anomalies seen in O supergiants are illustrated. HD 195592 shows spectrum of normal O9.7 Ib star while enhanced C III features and reduced N III and N IV features are exhibited by OC9.7Iab star HD 194280. The spectra of ON9.7 Iab stars BD+36° 4063 and HD 191781 have weak C III and enhanced N III features. ( Figure courtesy I.D. Howarth and Editorial Complutense S.A. Reproduced from *UV Astronomy: Stars from Birth to Death* with permission from Editorial Complutense, 2007).

## 2.2   O stars with C, N anomalies

The classification dichotomy for OB stars has been discussed by Walborn (1988, 2003). It is proposed that That OBC class corresponds to normal OB supergaints which already have CNO processed material mixed into their atmospheres. The OBN class displays even higher degree of this mixing. Walborn and Howarth (2000) have shown that the O stars also show ON, OC classes. In visual region, for example, N enhanced stars show weakening of C III $\lambda4650$ relative to N III $\lambda4640$ and the N III feature at $\lambda4097$ is weakened in C sequence as illustrated in Fig 3.

## 2.3   B stars

The spectra of B type stars are dominated by lines of He I and H I; former reaching a maximum strength near B2 while the later near A2. For normal B stars the ratios of lines of these two species are adequate to define the spectral types B0, B1.... B9, but the existence of He-strong and He-weak stars necessitate the usage of additional criteria such as Si IV $\lambda$4089/Si III $\lambda$4552 for early B and Si III $\lambda$4552/Si II $\lambda$4128-32 upto B3 and He I $\lambda$4471/ Mg II $\lambda$4481, He I/H I ratios are useful criteria for late B spectral types.

The luminosity criteria also depend upon the spectral types. Upto B2 the ratios of O II lines with H I or He I line is a sensitive luminosity indicator. The ratio He I $\lambda$4026/ O II $\lambda$4070,4076, He I $\lambda$4387/O II $\lambda$4416, H$\gamma$/O II $\lambda$4348 are also employed. Till B5, the ratios of Si III $\lambda$4552/He I $\lambda$4387 are useful, N II $\lambda$3995 feature and H I profiles can be used till late B.

In ultraviolet, accurate classification was achieved by Rountree and Sonneborn (1991) who preferred photospheric lines over the features arising in the stellar wind to maintain consistency with MK spectral types. In UV, the most useful spectral type criteria are the lines ratios Si II $\lambda$1264/ Si III $\lambda$ 1299, Si II $\lambda$1265 /Si III $\lambda$ 1341-1343, C II $\lambda$ $\lambda$ 1334,1335/C III $\lambda$ 1175,1176 and Al II $\lambda$ 1671/Al III $\lambda$1863. For luminosity calibration, Al III features $\lambda\lambda$ 1855, 1863 are useful for early B spectral types while Al II $\lambda$1671, Si II $\lambda$1265 and a host of Fe III lines at $\lambda\lambda$ 1891-1988 range are useful in late B types.

### 2.3.1   *B stars with helium anomalies*

A small number of early B stars exhibit very strong helium lines. The prototype of this group $\sigma$ Ori E has been classified B2 V h with "h" indicating helium strong. These stars show photometric and spectroscopic variations of the order of days. Bidelman (1965) had noticed that for HR 5378, the He I lines varied from above normal to below normal strength, while for other He-strong members the spectroscopic variations are between strong helium to normal helium strengths. Wade et al. (1997) noted variations in longitudinal magnetic field following a period similar to spectroscopic variations for HD 184927. These authors proposed a model with a patch of enhanced helium on the photosphere of HD 184927 near the positive magnetic pole. In this model, the inclination of magnetic axis with respect to rotation axis was bringing the enhanced helium patch in and out of view thereby explaining the observed photometric and spectroscopic variations. Similar models have also been used to explain the spectroscopic variations of magnetic Ap stars.

In He weak stars the helium lines remain weak and they have to be classified using the H I lines. These stars show unusual strengths of certain metallic lines and have been assigned subclasses accordingly. The Si stars have enhanced lines of Si II, similarly phosphorus-Gallium PGa, strontium-titanium SrTi stars and HgMn stars have very strong lines of corresponding elements.

### 2.3.2    *B stars with emission lines*

A Be star is believed to be a rapidly rotating star with a disk in the equatorial plane. In their spectra, the lower members of Balmer lines and sometimes even the lines of ionized species like Fe II are affected by emission components, the useful classification criteria are higher members of Balmer lines and Mg II feature at $\lambda 4481$. The Be activity ranges from mild (when the parent star can be classified) to extreme where the forest of emission lines prevent any reliable classification. Classification systems for emission line spectra have been proposed by Lesh (1968) and Jaschek et al. (1980). A high resolution atlas of emission and shell lines in Be stars has been published by Hanuschik et al. (1996). They propose a classification system for the emission spectra based upon inclination of the circumstellar disk, optical depth of lines and disk kinematics.

### 2.3.3    *B[e] stars*

B[e] stars are B stars that show the forbidden emission lines in their spectra such as those of [OI], [Fe II] near-mid IR excess suggestive of warm circumstellar matter around them. The classification criteria of B[e] class are given in Zickgraf(1998). Actually this definition covers a very diverse group of objects including pre-main sequence stars, symbiotic stars, planetary nebulae etc. Lamers et al. (1998) have proposed replacing the B[e] class with five classes. (i) "sgB[e] stars" which are luminous supergiants B[e] like MWC 300 in our Galaxy; (ii) "HAeB[e] stars" which are Pre-main sequence B[e] stars similar to Herbig AeBe stars; (iii) "cPNB[e] stars" which are like Pre-planetary nebulae with envelopes too compact for imaging but display emission line spectra similar to Planetary Nebulae; (iv) "SymB[e] stars" which show the emission lines of He I, FeII and absorption spectrum of TiO similar to M stars and (v) "UNCLB[e] stars" as the name suggests these are B[e] stars not fitting into any of the above classes.

A more extensive discussion with illustrative examples is given in the recent monograph by Gray & Corbally (2009).

## 2.4    A stars

The classification of A type stars is a challenging task. Nearly one third of A type stars exhibit different kinds of chemical peculiarities. Another source of difficulty is the large rotational velocities observed in a significant fraction of them not only broadens (and literally wash away the weak metallic lines), the local effective temperature and gravity vary over the surface of the rapidly rotating stars. The equatorial region will have lower temperature and gravity than the polar regions. The pole-on stars would appear systematically hotter and edge-on stars cooler. The well-known MK A0 standard Vega is a rapid rotator and had been known to be over-luminous by 0.7mag for its spectral type. Gray (1985,1988) had proposed that Vega may be a rapid rotator seen pole-on. This was later confirmed by Gulliver et al. (1991) from high resolution spectroscopy and by Aufdenberg (2006) through optical interferometry. A very detailed discussion of the rotation induced difficulties in classification can be found in Gray & Corbally (2009).

The spectrum of normal A stars (not easy to find) is dominated by hydrogen lines which are temperature as well as luminosity sensitive. The lines Ca II K, Ca I $\lambda4226$, Fe I lines at $\lambda\lambda$ 4045,4271 and 4383 are very good temperature and hence spectral type indicators.

The profiles of H I lines are very useful luminosity indicators till A7 while for late A types the Fe II and Ti II lines increase in strengths significantly from luminosity class V to I hence are very useful luminosity indicators. In ultraviolet the A-type spectrum becomes very crowded but Mg II h and k blend and C I feature at $\lambda1971$ are sensitive to temperatures while ratios C I $\lambda1657$/Al II $\lambda1670$ and blend at $\lambda1890$/C I $\lambda1931$ are sensitive luminosity indicators. In IR region the spectra of A stars are dominated by lines of Paschen and Brackett series. These lines show the same kind of temperature and luminosity sensitivity as Balmer lines in visual. The NIR Ca II triplet is sensitive to temperatures while O I $\lambda\lambda7771$-75 triplet, O I line at $\lambda8446$ and N I line at $\lambda8629$ are very sensitive to luminosity.

### 2.4.1  *Am stars*

Am stars or metallic-line stars are those A stars in which the spectral type determined from Ca II K line is earlier by at least five spectral types than that determined using metal lines. Although these objects were known to the members of Harvard classification team, the existence of this class was formally announced by Titus & Morgan (1940) who investigated a large sample of these stars in Hyades cluster. The class Am was included in the MKK atlas in 1943 using 63 Tau to define the Am characteristics. For this object the metallic line spectrum would give a spectral class of F2 III, hydrogen lines indicate A9 while Ca II K feature is similar to A2. Hence the assigned spectral class for 63 Tau is kA1.5hA9mF3III. Am phenomenon is attributed to the chemical separation driven by radiative and gravitational accelerations. This effect although present in all normal A stars, cannot be seen in rapid rotators since the meridional circulation for stars with $vsini > 90$ kms$^{-1}$ can completely mix the outer envelope wiping out the signature of chemical separation. Am stars being slow rotators the chemical separation overwhelms the meridional circulation. The elements like He tend to sink causing the disappearance of helium convection zone supposed to exist underneath hydrogen convection zone. The abundance peculiarities generated by chemical separation hence reach the outer surface making Am star.

### 2.4.2  *Ap stars*

Morgan(1933) had identified five groups of peculiar A type stars - Mn II, $\lambda$4200, Eu II, Cr II and Sr II stars. Later the feature at $\lambda$4200 was identified with Si II lines hence this group is now called Si II group. The MK spectral types for mild Ap stars can be easily determined with the help of hydrogen lines ; however for extreme Ap stars such as rapidly oscillating Ap stars (roAp stars) with distorted hydrogen profiles, the classification can be very approximate. In addition to these groups Ap stars contain groups with chemical peculiarities of different elements. Przybylski's star (Przybylski 1961) not only shows enhancement of lanthenides, but also that of Tc and $^{145}$Pm with half lives of 4.2 x $10^6$ and 17.7 yrs respectively pointing to *in situ* production of these elements as shown by Cowley & Bond. (2004). Strong magnetic field are of common occurrence and spotted surface model to explain Ap stars has been proposed. A comprehensive description of Ap phenomenon can be found in a review by Landstreet (2004) and Gray & Corbally (2009).

### 2.4.3  *$\lambda$ Boo stars*

The prototype of this group was included in MKK atlas where they described it as A0 stars with weak metal lines. These main sequence stars have broad spectral lines and their spectra appear similar to A-F stars with weak metal-lines and Mg II $\lambda$4481 / Fe I$\lambda$4383 ratio is smaller than one. However, C,N,O lines appear of normal strengths. These stars are also given 3 spectral types based upon hydrogen lines, Ca II K line and general metal-line spectrum. The spectral type for $\lambda$ Boo is A3Vak B9.5m B9.5$\lambda$ Boo . Paucity of metal-lines in these population I objects is quite unusual. The observed abundance pattern was explained by Venn & Lambert (1990) in terms of dust-gas separation in the circumstellar matter near the star where the condensable elements are locked into the grains. The metal-poor gas is accreted by star. According to Turcotte (2002) an accretion rate of $10^{-11} M_\odot yr^{-1}$ is required to prevent meridional circulation from canceling the effect of accreted metal-poor gas and hence retaining the the observed abundance anomalies.

## 3.  F stars

The spectra of F stars are delightful with good representations of atomic species without getting unduly crowded. Molecular bands of CH can be seen from F5 onwards. These stars are important tools to study young disk, thick disk as well as halo populations.

Hydrogen lines serve as prime indicators of spectral types, although metal lines such as Fe I $\lambda\lambda$ 4046, 4383, Ca I line at $\lambda$4226 grows rapidly in later spectral types. From F5 to later spectral types, ratios Fe I $\lambda$ 4046/H$\delta$, Ca I $\lambda$4226 / H$\delta$, H$\gamma$, and Fe I $\lambda$4383/H$\gamma$, are sensitive indicators. However for metal-poor stars they will give, earlier types than actual spectral types. Hence the hydrogen lines are the most reliable indicators.

The ionized lines of Fe and Ti at $\lambda\lambda$ 4172–8, $\lambda\lambda$ 4395–4400, $\lambda$4417, $\lambda$4444 have strong luminosity dependence hence their ratios with features that are not luminosity sensitive e.g. Fe I $\lambda$4046, $\lambda$4271, $\lambda$4383 and Ca I line at $\lambda$4226 are useful. For F6 and later spectral types (through G ) upto K types, the ratios of Sr II $\lambda \lambda$ (4077, 4215) with Fe I $\lambda$4046 or H$\delta$ are very sensitive luminosity indicators. At 8400-8750 Å  spectral region, Paschen P16-P12 and Ca II lines at $\lambda\lambda$ 8498, 8542 and 8662, O I $\lambda$8446 triplet show strong luminosity effects.

### 3.1   Metal-poor F type stars

While classifying F stars, N.Roman (1950, 1952, 1954) had identified metal-weak stars which also had high radial velocities and connection between population and metallicity was established. These studies have also helped in refining the Baade's population scheme by including an intermediate population II which consists of objects with kinematics, metallicities, and concentration towards Galactic plane, intermediate to extreme population II and thin disk population. This intermediate population of stars are now called thick disk stars.

The early attempts to classify metal-weak stars used hydrogen lines for temperature classification and metallicity types were assigned by identifying the earlier MK standards with the same strength of metal lines as program stars. This system was far from satisfactory since the spectral pattern is a function of temperature and match of metallic line spectrum of early spectral type MK standard with metal-weak star was never satisfactory.

A more accurate approach is proposed by Gray (1989) who defined sequences of metal-weak standards which run parallel to MK standards (of solar metallicity) which were assigned metallicity class 0. The metal-poor sequences are labeled $-1$ and $-2$ with the later being the most metal-poor star of intermediate population. The temperature types are again assigned based upon hydrogen lines but metallicity classes are assigned based upon metallicity independent Cr I / Fe I ratio using Cr I resonance triplet at $\lambda\lambda$ 4254, 4275, 4290 and Fe I subordinate lines at $\lambda\lambda$ 4250, 4260 and 4326. Gray also proposed to set up a preliminary sequence for metallicity classes $-3$, $-4$, $-5$ thereby extending the scheme to F type extreme population II stars. The metallicity classes of Gray are well correlated with the Strömgren metallicity index *m1*. The major hurdle in the implementation of this scheme is the need of a large number of metal-poor standard stars over the whole range of metallicities.

## 4.   GK stars

They constitute a family of well-studied stars. Their spectra are rich in atomic lines but molecular lines become more prominent in K stars. The emission lines are present in GK dwarfs due to chromospheric activities which are also seen in pre-main sequence stars as well as in supergiants losing mass in the late stages of their evolutions. Giants and supergiants also exhibit chemical peculiarities either caused by mixing of nucleosynthesis products to the surface or through material transferred from the evolved binary companion.

## 4.1 Spectral type criteria

The molecular band of CH near 4300Å referred as G band and Ca I $\lambda$4226 become stronger towards later G subclasses. The Ca I $\lambda$4226 feature becomes very strong at K5. Although ratios such as Fe I $\lambda$4045/H$_\delta$, Fe I $\lambda$4144/H$_\delta$, Fe I $\lambda$4383/H$\gamma$ are still useful, Mg I lines at $\lambda\lambda$ 5167, 72, 83 become stronger towards later spectral type (this feature is also sensitive to luminosity). At K5 the MgH bands are good spectral type indicators. For metal-poor GK stars, ratios Cr I $\lambda$4254/Fe I$\lambda$ 4250, 4260 and Cr I $\lambda$4290 / Fe I$\lambda$ 4326 can be used. Near infrared region contains lines of Fe I at $\lambda$8468, 8514, Ti I line at $\lambda$8435 which increase towards later spectral types. K band region (around 2$\mu$m has many good spectral type indicators such as ratios of Brackett $\gamma$ line at 2.166 2$\mu$m with Na I and Ca I features nearby for G type, while CO band at 1.62 2$\mu$m is useful from K5 to M5 spectral types. For oxygen-rich late K and M giants, SiO and P-branch of OH ( in L band region) are sensitive to spectral types.

## 4.2 Luminosity type criteria

In optical region, the ratios of Sr II $\lambda$4077/Fe I $\lambda$ $\lambda$4046, 4064, 4072 and CN bands with band heads at $\lambda$4215, 3883 are sensitive to luminosity for G stars. For spectral types K5 and later the ratios of MgH/TiO blend near $\lambda$4470 is very sensitive to luminosity. Near infrared Ca II triplet at $\lambda$ $\lambda$ 8498, 8542, and 8662 are very sensitive to luminosity at K0. For later spectral types ratios of CO bands at 1.62 $\mu$m and 2.29 $\mu$m with Na I, Ca I, Mg I lines in 2 $\mu$m region are very sensitive luminosity indicators.

## 4.3 Sub-groups exhibiting emission features

Chromospheric activities in the sun and sun-like stars have been studied in detail. FGK stars of luminosity class IV and V contains active stars such as RSCVn binaries which have emissions in their Ca II H & K and H$\alpha$ lines. They also exhibit strong emission in UV X ray and radio wavelengths indicative of active coronae.

T Tauri stars which are precursors of solar type stars are characterized by emission features in H, Ca II, Fe II and Na I lines. The Li I $\lambda$6708 is very strong.

**4.4   GK evolved stars**

GK giants contain groups of stars showing various kinds of chemical peculiarities such as strong CN stars, weak G band stars and Barium stars. In weak G band stars the mixing of CN processed material (with reduced carbon and enhanced nitrogen) is considered responsible for the weak G band of CH, while Barium stars are carbon enriched and show strong lines of Ba and other s-process elements due to the material transferred from a AGB companion.

GK supergiants also contain pulsating stars such as classical Cepheid, RV Tauri stars and W Vir Variables (population II Cepheids). Each group follows its own period-luminosity relation and therefore are useful as distance indicators. W Vir and a few RV Tau stars show phase dependent emission components in Hydrogen, Ca II and Na D lines of their spectra. Abundance peculiarities caused by mixing of CN processed material as well as those caused by selective removal of condensable elements are seen in some RV Tauri stars as well as in W vir stars as summarized by Giridhar et al. (2005) and Maas et al. (2007).

# 5.   M, S, C stars

These stars from a parallel sequence where at similar temperatures spectra appear very different since different molecules of C or O dominate in their spectra.

**5.1   M stars**

M stars have the distinction of exhibiting very large range in ages and radii. The spectra of these cool objects are very crowded with strong atomic as well as molecular lines making the continuum normalization impractical. Hence normalized fluxes are plotted as a function of wavelength.

The ratio of Ca I $\lambda4226$/Fe I $\lambda4383$ can still be used as spectral type defining criterion upto early M, but TiO takes over from M5, and VO band are useful at M9. The resonance lines of K I at 7665 and 7699 Å  are both temperature and luminosity sensitive. For metal-poor stars, abundance independent spectral type criteria like ratios of TiO bands $\lambda$ 4808/TiO $\lambda4955$ can be used.

Ca I$\lambda4226$ are strong in dwarfs but become weaker in giants and supergiants. On the other hand, ratios of $\lambda5250$/$\lambda5259$ blend is strongest for supergiants and reduces towards dwarfs due to the fact that Fe I $\lambda5250$ is stronger as luminosity increases. The shape of MgH/TiO blend also changes drastically due to weakening of MgH component at high luminosity.

### 5.1.1   *Mira Variables*

These are very interesting M type variables showing long period (80-1000days) and large amplitude (2.5 to 10 mag.) light variations. The spectra contain Ti O bands similar to those of M giants but emission lines of hydrogen, Fe II, Ti I and Fe I are seen. The strength of these features vary with luminosity cycle. The weak Fe I emission at $\lambda\lambda$ 4202, 4308 is probably produced by fluorescence; the upper layer being populated by UV photons caused by shocks. However, the metallic line emission varies independently of Hydrogen lines. At certain phases all these emissions are weak or absent and the spectrum would look like that of normal M giant. Another interesting feature is that spectra show veiling effect (weakening of lines at certain wavelengths while at other wavelengths the lines of the same specie remains unaffected) which varies from cycle to cycle. The AlO band at $\lambda4842$ varies from strong absorption in the cycle with least veiling to emission feature in cycles with large veiling effect. Mira variables are believed to be surrounded by extended atmospheres which experiences pulsation induced shocks.

## 5.2   Carbon stars

Carbon stars were discovered by A. Secchi who was able to identify the molecular bands in the spectra with carbon molecules. In HD classification they were subclassified into R and N branches. The N type was further subdivided into Na, Nb, Nc on the basis of redness of their energy distribution. Nc has no flux shortward of $H_\beta$. R stars have strong carbon bands but more continuum flux in blue-violet compared to N stars. The R subclasses were defined in the order of decreasing violet continuum fluxes. There is a certain amount overlapping properties between R8 and Na.

Keenan & Morgan(1941) attempted a reclassification of all carbon stars into a single temperature sequence and adding $C_2$ strength index which indicated the strength of Swan band. However, this C0, C1, .. system was not found satisfactory as the chosen temperature criteria like Na D strength were affected by overlapping molecular bands. Also it was realized that R and N belonged to different populations.

A revised *MK Carbon star classification system* was introduced by Keenan (1993) which divided C stars into five groups.

### 5.2.1   *The C-R stars*

These are warmest of carbon stars characterized by strong blue-violet continuum flux and have temperatures similar to G4 to M2 stars. The spectra contain prominent CN bands at $\lambda4215$ and $C_2$ Swan band at $\lambda4237$ and $\lambda5165$.

### 5.2.2    *The C-N stars*

In these stars the continuum flux is shifted to longer wavelength compared to C-R stars and there is no flux shortward of 4400Å . The $C_2$ bands are weaker than in C-R stars but lines of s-process are more enhanced. The spectra otherwise could look similar to those of G6-M6 stars. Compared to C-R stars, they are more concentrated in galactic plane.

### 5.2.3    *The C-J stars*

These stars are characterized by a large abundance of isotopic $^{13}$C. In these stars $^{12}C/^{13}C$ ratio of about 10 has been observed while the solar system value of this ratio is about 90. This quantity can be measured through ratios of $^{13}$CN $\lambda6260/$ $^{12}$CN$\lambda6206$ and $^{12}C^{13}$C $\lambda6168/$ $^{12}C^{13}$C $\lambda6122$. The lines of s-process elements are similar in strengths to those of C-R stars.

### 5.2.4    *The C-H stars*

The bands of CH are very strong; the strength of P-branches of CH band is so strong that it is visible as longward depression. These objects also show s-process enhancement and a wide range in $^{12}C/^{13}C$ ratio.

### 5.2.5    *The C-H$_D$ stars*

These are hydrogen deficient carbon stars and can be identified through almost complete absence of G band of CH lines, and lines of hydrogen are almost absent. However, lines of CN and $C_2$ are anomalously strong. This group contains R CrB variable which shows drastic light decline as well as exotic spectral variations.

## 5.3    S stars

The spectra are characterized by bands of ZrO although other molecular bands of Ti O, YO and LaO are also seen. Many S-stars are long period variables. Based upon molecular dissociation calculations, Scalo and Ross( 1976) demonstrated that sequence $M - MS - S - SC - C$ represents an increase in the ratio of carbon to oxygen. This ratio is about 1 in SC stars.

**5.4   Symbiotic stars**

The spectra of these stars contain features seen in cool giants as well as those seen in planetary nebulae; well known member of this class is Z And. The symbiotic stars (SYSs) are now recognized as interactive binaries with large orbital separation and mass accretion onto the hot component play a very important role in controlling the basic parameters as well as evolution of symbiotic stars. These objects show large variations in brightness (upto 3 mag. in U) as well as in spectral appearances. A summary of their properties can be found in Mikolajewski (2003).

In UV the spectrum is dominated by the hot component while optical spectra are contributed mostly by cool (G-K) primary component. However, the spectrum contains strong emission in the hydrogen lines and highly ionized species normally associated with planetary nebulae such as He II, [O III], [Na III], [Fe V], C III and N III.

Symbiotics are classified into two types. The yellow s-type SYSs with G or K primaries have low metallicities and high space velocities similar to Galactic halo objects. The red s-type SYSs have M giant primaries and show Galactic bulge characteristics. Fig. 4 contains the spectra of a few red s-type SYSs stars.

# 6.   M and L Dwarfs

The last two decades have witnessed a spectacular progress in the classification of the objects at the cool end of spectral sequence largely due to increased detector sensitivity in IR. Although M supergiants were known from the beginning, M dwarfs could be identified only when the precision parallaxes of faint M stars showed them to be nearby objects. It became apparent that they were intrinsically low luminosity objects. Initial classification of their spectra had been done by Adams et al. (1926).

The spectral types using spectroscopic standards were defined by Boeshaar(1976) upto M6 and this work was further extended upto spectral type M9 by Boeshaar & Tyson (1985). The classification has been done using the strength of Ti O bands for early M dwarfs and using the ratio of VO at $\lambda 5736$/TiO $\lambda 5759$ and using the strength of CaOH at $\lambda 5330$ for late M dwarfs. A more complete and precise classification of M dwarfs has been carried out by Kirkpatrick et al. (1991) using the extended spectral coverage upto 9000Å . These authors showed that in 5000-9000Å spectral regions there is a noticeable strengthening of TiO bands from mid K spectral type to M6. For later spectral types VO bands alongwith the increasing slope of the pseudo-continuum at long wavelengths are excellent spectral type indicators. A list of primary standards for M dwarfs has been given in Table 9.1 of Gray & Corbally (2009).

The IR spectrum of M dwarfs contain neutral atomic lines of Ti, Ca, Fe, Si, Mg, Al, Na, K and Mn alongwith molecular lines of FeH, CO and $H_2O$. In early M the atomic features dominate while molecular features become prominent from mid-M onwards. M dwarfs could be distinguished from M giants from the strengths of Na I, K I and FeH features at near IR which weaken in giants. Metal-poor M dwarfs show weakening of TiO bands while hydride bands are much less affected. Metal-poor M dwarfs are referred as subdwarfs (sd) for historical reasons although the term is misleading. Similarly, extremely metal-poor dwarfs are called extreme subdwarfs (esd).

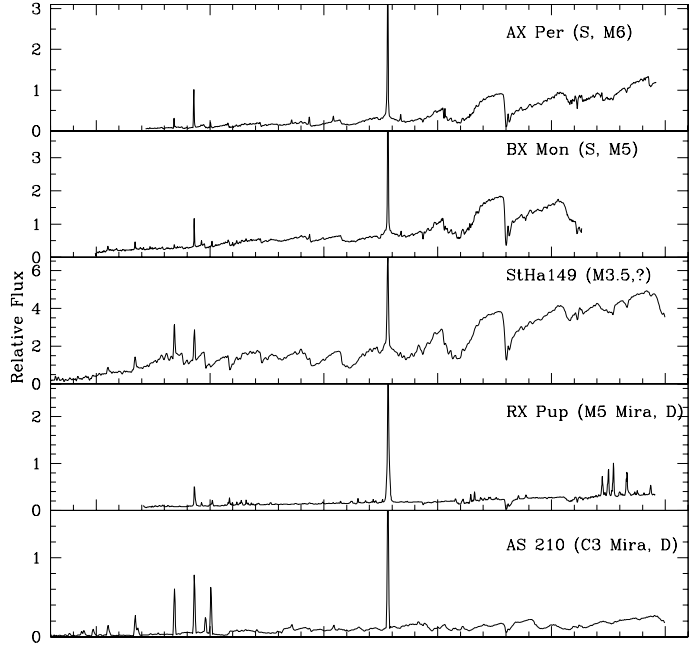

**Figure 4.** The representative spectra of a few red s-type SYSs symbiotic stars obtained from VBO observatory, Kavalur. Figure courtesy G.C. Anupama.

## 6.1 Brown Dwarfs

These objects are still in the stage of gravitational contraction and do not have hydrogen fusion at their core. They are less massive than M dwarfs but the spectra appear intermediate between M-type dwarfs and M-type giants. Notwithstanding the lower mass they have the same temperature range of M stars since they did not have sufficient time to cool and have larger radii as they are still contracting. In a sample of brown dwarfs covering a large range in ages, gradual weakening of Na, K lines and FeH, CaH bands from oldest to youngest brown dwarfs have been observed. At infrared wavelengths M dwarfs can be easily identified through Na, K lines and hydride bands while brown dwarfs exhibit much stronger $H_2O$ features and weaker CO features.

## 6.2    L Dwarfs

The spectrum of early L dwarfs is crowded due to lines of neutral alkali elements such as Na I, K I, Rb I, Cs I and sometimes those of Li I, TiO, VO, CrH, FeH and CaOH. Towards mid L, the resonance lines of Na I, K I become tremendously strong and hydride lines are also strengthened while oxide bands almost disappear. In late L, alkali features continue to remain strong while hydride bands weaken. Fig. 5 gives the identifications of prominent features observed in M9 V, L3 V and L8 V class objects.

A comprehensive study of L dwarfs has been done by D.Kirkpatrick and his collaborators. Optical classification along with a set of reference stars to define L spectral types is given in Kirkpatrick et al. (1999). Fig. 6 illustrates the lines and bands seen in M 7 to L8 dwarfs . The optical L dwarf sequence has been defined till L8 beyond which there is an abrupt change in spectral appearance.

Although L type stars do not have luminosity types like giants and supergiants, the observed variations of gravity sensitive features show that the known sample of L dwarfs displays a range in gravity. One could differentiate a low gravity L dwarf from M dwarfs through weakness of TiO and the presence of VO band, but the lines Na I, K I and hydride bands would be weaker than L dwarfs indicating lower gravity.

At infrared wavelengths, in addition to $H_2O$ bands (which are stronger than those in M dwarfs), $CH_4$ band at $3.3\mu m$ makes it appearance. At near infrared the ultra strong K I absorption at $0.77\mu m$ reduces the flux at the core of the line by a factor of 1000.

The temperature of M dwarfs have been estimated by López-Morales (2007) and for L and T dwarfs by Vrba et al. (2004) using empirical determinations of radii and luminosities. The temperatures approximately range from 3800K to 3000K for M0 -M5 dwarfs and 2400K-1300K for M9 to L8 dwarfs. For spectral types earlier than M dwarfs, we have seen the strong dependence of spectral appearance on temperatures. As the temperature is reduced the ionised species weaken, neutral atomic lines strengthen and at later spectral types (lower temperatures) molecular bands dominate. From M9 to L types there are drastic changes over the relatively small temperature span. Here the formation of condensates cause disappearance of TiO and VO molecular bands. Chemical equilibrium calculations by Lodders (1999), Burrow & Sharp(1999), Lodders(2002) have explained the the locking of Ti, V and other refractory elements into dust grains as different complex compounds. The removal of these elements leaves only the less refractory neutral alkali metals as the major sources of opacity. This explains the great strengths of Na I, K I lines with wings extending to several hundred Å in the spectra of late L dwarfs.

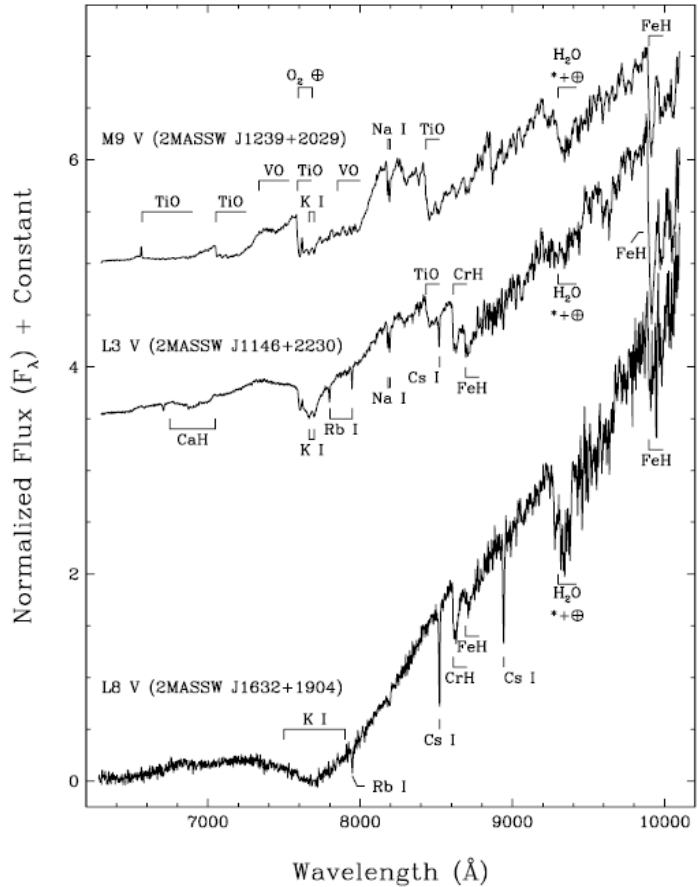

**Figure 5.** The identifications of spectral features showing large variations from M9 V to L8 V class is presented. (Figure courtesy Kirkpatrick et al. 1999. Reproduced from ApJ, 519, 802 by permission of the AAS).

## 6.3 T dwarfs

These objects can be distinguished from L dwarfs by the presence of $CH_4$ absorption in IR region, stronger $H_2O$ and the presence of $NH_3$ bands. The first T dwarf Gliese 229B identified by Nakajima et al. (1995) and Oppenheimer et al. (1995) as low luminosity companion to M1 star Gliese 229. The latter happens to be M dwarf spectral standard (Kirkpatrick et al. 1991). Fig. 7 shows the spectra of L6 and L8 dwarfs alongwith that of Gliese 229B. The T dwarf spectrum has strong $CH_4$ band at $2.4\mu m$. With close resemblance of their spectra to those of planetary spectra, T dwarfs can be considered as objects intermediate to low mass stars and planets. An excellent review of developments

    *Sunetra Giridhar*

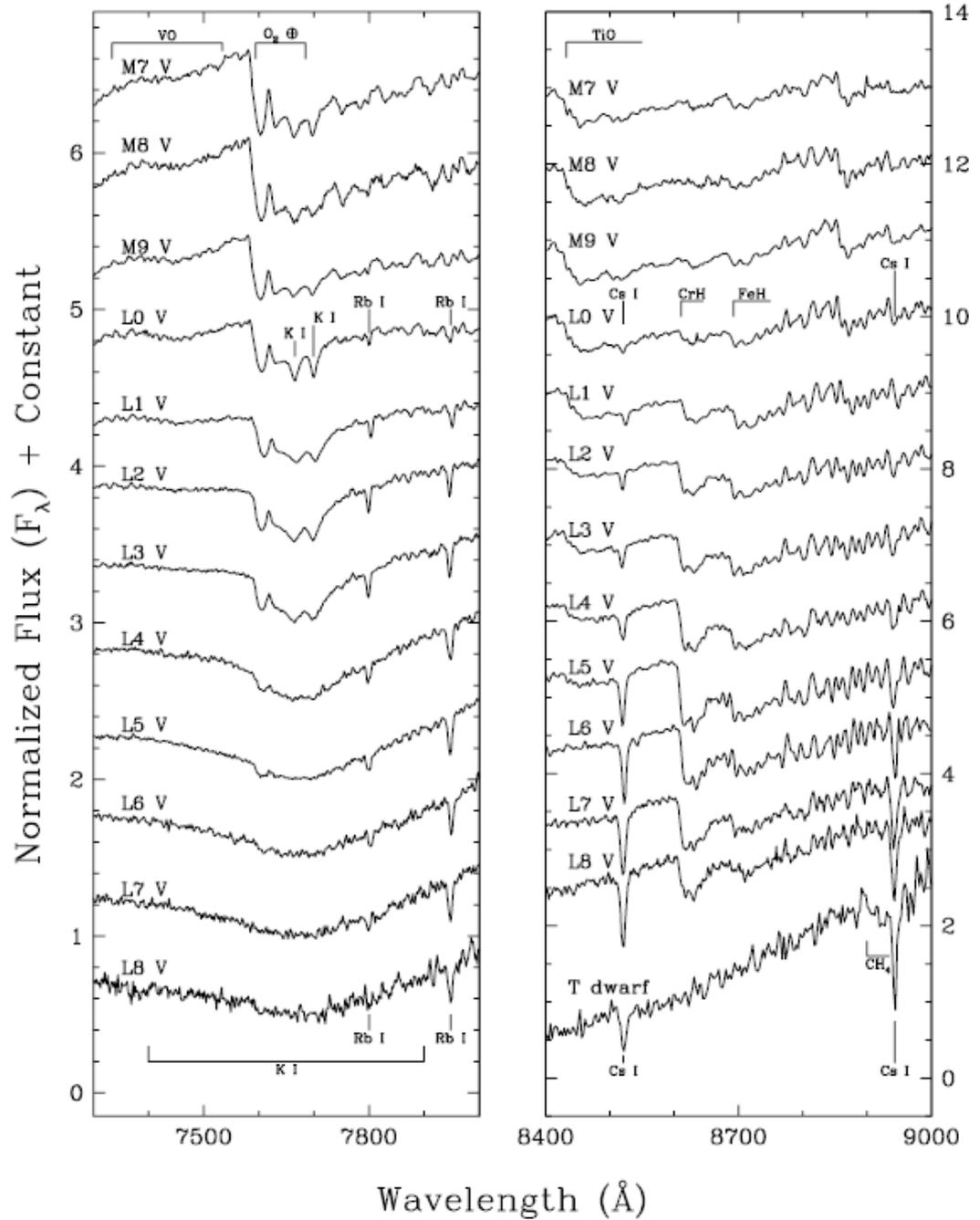

**Figure 6.** The spectral variations seen over spectral types M7 V to L8 V. (Figure courtesy Kirkpatrick et al. 1999. Reproduced from ApJ, 519, 802 by permission of the AAS).

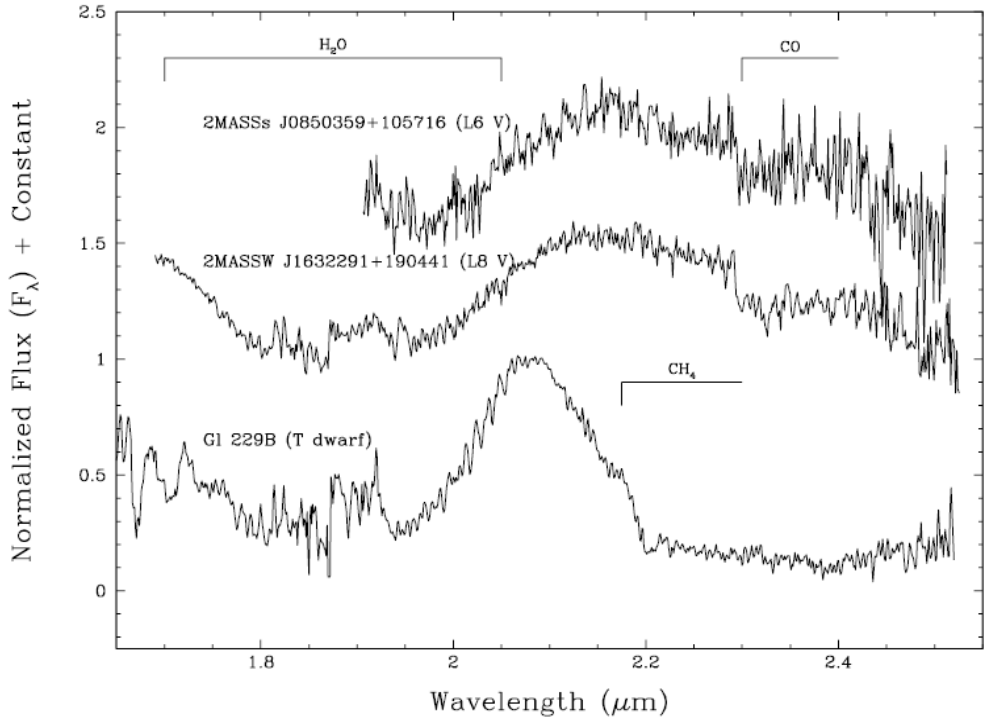

**Figure 7.** The spectra of two late L dwarfs with Gliese 229B, the first identified T dwarf. (Figure courtesy Kirkpatrick et al. 1999. Reproduced from ApJ, 519, 802 by permission of the AAS).

in this field including the discoveries through various surveys can be found in the chapter on T dwarfs by Adam Burgasser in the monograph by Gray & Corbally (2009).

The spectra of T dwarfs contain atomic features such as Na I D lines, $H_\alpha$ emission, K I lines at $\lambda\lambda$ 7665, 7699 and IR lines at 1.169 and 1.252 $\mu$m, Rb I lines at $\lambda\lambda$ 7800, 7948 and Cs I lines at $\lambda\lambda$ 8521 and 8943. In optical and red spectral regions molecular features of CaH, CrH and FeH are identified at 0.675, 0.8611, 0.8692 $\mu$m, respectively. Near IR and IR region contains several features of $H_2O$, CO, $CH_4$ and $NH_3$ and collisional induced $H_2$ absorption (CIA $H_2$) features at 1.8 to 2.8$\mu$m.

The spectral classification of T dwarfs using IR spectra in 1.0 to 2.5 $\mu$m was undertaken in parallel by Burgasser et al. (2002) and Geballe et al. (2002); both schemes employing the strengths of $H_2O$ and $CH_4$ band strengths as classification criteria. Notwithstanding the inhomogeneous distribution of the sample and non-uniform resolution of the spectra used, Burgasser et al. systematically arranged spectra of 24 dwarfs and selected a sequence of seven subgroups representing even spectral variations and assigned them subtypes T1 through T8. On the other hand, the data set used by Geballe et al. (2002) had better and uniform resolution and homogeneous distribution of sample stars over the spectral subtypes. This scheme used the spectral indices sampling the $H_2O$ and $CH_4$ bands at different wavelengths. The usage of $CH_4$ band strengths at 1.6 and $2.2\mu m$ enabled Geballe et al. (2002) to distinguish early T dwarfs from late L dwarfs which had only $2.2\mu m$ feature of $CH_4$ but not $1.6\mu m$ one. This scheme however did not present standards to represent these subtypes which Burgasser et al. (2002) had done. A collaboration of these two groups has resulted in a unified classification scheme similar to MK process. It is presented in Burgasser et al. (2006) which is now considered primary method of T dwarf classification. A large number of primary and secondary standards for each subtype are presented. The classification of unclassified T dwarf can be done by comparing the IR spectrum with those of spectral standards. Alternately, one can use five spectral indices which are ratios of the average flux densities measured at different wavelengths essentially measuring the strengths of $H_2O$ features at 1.15 and $1.4\mu m$ and those of $CH_4$ features at 1.32, 1.65 and $2.2\mu m$.

Although T dwarfs are brightest at 1.2 to 1.35 $\mu m$ and 1.0 to 2.25 $\mu m$ region offers excellent classification criteria, 0.6 to 1.0 $\mu m$ region has also been explored and found to contain features such as Na I, K I lines with deep and pressure broadened wings affecting the slope of the red continuum, red and near infrared colours of T dwarfs. Features of Rb I, Cs I, FeH band at $\lambda 9896$ and $H_2O$ absorption in $\lambda$ 9250-9800 region show considerable variations over the spectral types. Burgasser et al. (2003) developed a classification scheme using red-optical spectra which was tied to the IR spectral standards. The optical spectral types inferred for these objects by Burgasser et al. (2003) were generally consistent with IR classification within 1 subtype uncertainty . It is therefore proposed that optical and IR spectral morphologies are well correlated unlike L dwarf regime where optical and IR subtypes show large differences largely attributed to condensate formation effect.

Mid infrared region (6-10 $\mu m$) explored by Cushing et al. (2006) also contains features of $H_2O$, $CH_4$ and $NH_3$ which are sensitive to spectral types and has been found very useful.

The trend in the temperatures of T dwarfs as a function of spectral type has been studied by Golimowski et al. (2004). The temperatures are based on luminosity determinations from parallax and broad-band photometric measurements (Legget et al. 2002, Vrba et al. 2004). The spectral types for T dwarfs later than T4 appear to be correlated with temperatures (the temperatures being approximately 1000K to 700K for T4 to T8 dwarfs). The early type T dwarfs however have nearly the same temperatures as the late type L dwarfs. A similar flattening in temperature versus bolometric luminosity relation has also been observed. It could be related to evolution of condensate clouds across L/T transitions. We have seen that condensates strongly influence the IR spectra of L dwarfs and early type T dwarfs, but appear to be largely absent in mid- and late type T dwarfs. It is proposed by Ackerman & Marley (2001), Burgasser et al. (2002) Tsuji & Nakajima (2003) that the depletion of condensates could play a significant role in the transition between the L- and T dwarf classes. The mechanism for this depletion is not fully understood but many factors such as surface gravity, metallicity, rotation in addition to temperature may affect the process.

Two objects cooler than T8 dwarfs have been detected through UKIDSS and CFHTLS survey (Warren et al. 2007; Delrome et al. 2008). They exhibit $NH_3$ feature in the NIR spectra at 1.55 $\mu$m. Although they could mark the beginning of a new cooler class "Y dwarfs" proposed by Kirkpatrick (2005), they could also be considered as T dwarfs later than T8, there is no consensus on this issue.

## 7. White Dwarfs

White dwarfs are the final stage of evolution of medium and less massive stars when they have exhausted all their fuel, shed their atmospheres leaving behind the dense CO core.

The first white dwarf (WD) discovered by W. Flemming and E.C. Pickering in 1910 was 40 Eri B. They were puzzled by finding this A-type star among low luminosity objects. It was followed by the study of spectrum of the faint companion of Sirius by W. Adams (1914). S. Eddington (1924) recognized these objects as very dense compact ones and christened them "white dwarfs". Subsequently, Fowler (1926) showed that the degenerate electron gas pressure could support such dense objects.

The spectra of WDs were found to be dominated by broad lines of hydrogen and helium; lines of other elements were also detected in some subclasses. With a very limited sample of stars, Kuiper (1941) attempted classification based upon the lines of dominant specie present like A for hydrogen lines, F for Ca II lines, G for weak metallic line blends and "con" for a continuous featureless spectrum. The prefix "D" was chosen by Luyten (1952) who also proposed numeric subtypes to indicate the temperatures. This work was further refined by Greenstein (1960) who used colour index and the strength of H$\gamma$ as temperature indicator. With the advent of better detectors larger database of higher resolution spectra could be obtained; it was appreciated that these objects had a very large range in temperature like $T_{eff} > 50,000$K to $< 5000$K. A modern classification scheme was initiated by Sion et al. (1983). The refined version of it has been in use presently is described in Liebert & Sion (1994). This scheme has the following primary spectroscopic symbols characterizing various subtypes listed below.

DO: Spectrum dominated by He II, H I and He I may be present

DA: Hydrogen lines are dominant features

DB: Spectrum dominated by He I lines

DC: Essentially featureless continuum

DZ: Metal lines are strongest features

DQ: Carbon lines either atomic or molecular are dominant features

other secondary symbols used

P - Magnetic WD with detectable polarization

H - Magnetic WD without detectable polarization

X Peculiar or unclassified spectrum

? uncertain classification

E emission lines present

V denoted variability

Following these primary and secondary symbols the temperature indices from $0-9$ were used in Sion et al. (1983) scheme; $\theta = 50400/ T_{eff}$ gives an integer which would be temperature indicator like DA1, DA2...... DA9 in descending temperature order.

With increased accuracy of WD temperatures estimates (from the use of more sophisticated non-LTE synthetic spectra codes and atomic data) and detection of very hot WDs, the temperature definition in the scheme of Sion et al. (1983) was found unsatisfactory as it does not provide necessary temperature discrimination towards the hotter end of WDs. A refinement described in Liebert and Sion (1994) involves for the hottest stars, assignment of temperature index of decimal 9 to decimal 1 (.9 to .1 avoiding zero prefix to prevent confusion with the hot primary spectral type of O). For stars with $T_{eff}$ of 200,000 K, the index would be .25. The use of such decimal half integer temperature allows for refined temperature classification. Thus a sequence of DA stars may extend from DA.25, DA.5, DA1, DA1.5.... DA13.

In addition to the temperatures, better accuracy in gravity determinations by fitting all the of Balmer lines, it is apparent that WDs have a large gravity range (log g) $7 < > 9$. The optional symbols "d" and "n" are used to indicate diffuse features in high gravity and narrow features in low gravity objects.

An interesting finding in the recent times is that of a group of cool DAZ WDs showing the evidence of debris disk (von Hippel 2007). The existence of dusty disk is inferred from the IR excess after $2\mu$m. The accretion of refractory elements by the WD probably accounts for the lines of Ca, Mg and Fe in their spectra.

Yet another type of hot DQ star has been found by Dufour et al. (2007) which shows the lines of ionized carbon (C II). Hitherto known DQ WDs only exhibited $C_2$ features and had only traces of carbon in their atmospheres. Dufour et al. (2007) have done evolutionary calculations and shown that hot DQ stars have dredged up carbon from the core and their atmosphere can have high abundances of carbon. A few hot DQ stars also have detectable O I and O II lines.

Detection of these new kinds of WDs offer new challenges to the evolutionary models of WDs.

## 8.   Automated methods of classification

With the availability of multi-object spectrometers on many world class telescopes several large scale stellar surveys have been undertaken and planned weing the ground based as well as space missions like GAIA, ASTROSTAT. To handle large volumes of data it is necessary that automated procedures are used. These procedures would also have the advantage of objectivity and homogeneous data sets would be generated which is essential for developing models of Galactic structure and evolution. Automated procedures would also help in quick identification of peculiar objects.

The semi-automated and fully automated procedures have been developed in past and there is considerable progress made in the last decade. Most commonly used methods are based upon (a) Minimum Distance Method (MDM), (b) Principal Component Analysis (PCA) and (c) Artificial Neural Network (ANN). These methods are described in an excellent review paper by Bailer-Jones (2002).

The MEM based on metric distance technique is described in Kurtz (1984) and LaSala (1994). Here the metric distance between program star and standard star is to be minimized. Vansevicius and Bridzius (1994) used MDM with $\chi^2$ weighing to estimate SpT and $M_V$ from Vilnius photometric indices. An accuracy of 0.7 was achieved for SpT and 0.8 mag for $M_V$ over spectral type range O5 to M5. Malyuto (2002) has applied improvised version of this method by adding maximum likelihood procedure to simulated photometric data for classification of stars.

PCA is a method of representing a set of N dimensional data by means of their projection onto a set of optimally defined axes. These compressed data sets are used as input for neural networks. Bailer-Jones et al. (1998) and Singh et al. (1998) have demonstrated that precise calibration could be done using these compressed spectra and that the optimal compression also results in noise removal. Singh et al. (2006) have used a variation of PCA technique to restore missing data in a sample of 300 stars in Indo-US coudéfeed spectral library.

Neural network is a computational method which can provide non-linear parametrized mapping between an input vector (a spectrum for example) and one or more outputs like SpT, LC or $T_{eff}$, log $g$ and [M/H]. The method is generally supervised, it means that for the network to give required input-output mapping, it must be trained with the help of representative data patterns. These are stellar spectra for which classification or stellar parameters are well determined. The training proceeds by optimizing the network parameters (weights) to give minimum classification error. Once the network is trained the weights are fixed, the network can be used to produce output SpT, LC or $T_{eff}$, $\log g$ and [M/H] for an unclassified spectrum. The ANN has been used in very large number of stellar applications.

Gulati et al. (1994a) used a multilevel tree structured neural network configuration to classify low resolution IUE spectra with an accuracy of one subclass for the most spectra. Vieira and Ponz (1995) have used ANN on low-resolution IUE spectra and have determined SpT with an accuracy of 1.1 subclass. Although they attempted classification also with MDM, the errors of classification were larger than that of ANN. More recently, ANN based tools have been developed by Bora et al. (2008) for classifying a large number of stars from IUE database with an accuracy of 3-4 spectral subtypes and have also estimated colour excesses for these stars with an accuracy of 0.1mag using the simulated photometric bands data defined for TAUVEX mission. Bora et al. (2009) have developed an ANN based method for automated segregation of stars from the galaxies using UVBLUE library and IUE low resolution spectra.

In visual region Gulati et al. (1994b) have used ANN based approach on the database of 158 digitized spectra and have classified these stars with an accuracy of 2 spectral subclasses. Bailer-Jones, Irwin and von Hippel (1998) used ANN to classify spectra from Michigan Spectral Survey with an accuracy of 1.09 SpT. Visual-near IR spectra were classified by Weaver & Torres-Dodgen (1997) using a two step approach. At first a coarse classification is being done to get main spectral class say F, then it is further classified by more specialist network for that class. This approach results in an accuracy of 0.4 to 0.8 for SpT and 0.2 to 0.4 in LC.

The ANN scheme has been used by Gupta et al. (2004) to classify a large sample of bright IRAS sources using Calgary database of IRAS spectra in 8-23 $\mu$m in 17 predefined classes.

A more recent review of these automated methods and their applications to different surveys has been summarized by Giridhar, Muneer & Goswami (2006).

## 8.1  A New Spectral class Encoding system

A team comprising of Myron Smith, Richard Gray, Christopher Corbally, Randall Thompson, and Inga Kamp has developed a new spectral class encoding system which allows the users to obtain information about selected groups of stars from data archives like VO, MAST etc. The procedure as given by these authors to identify all stars of a chosen spectral type and luminosity class is as follows. The team has designed a "spectral class" nomenclature system for spectra of stars across the HR Diagram into a finite number of bins. The nomenclature scheme has the form TT.tt.LL.PPPP, where TT and tt are numerical digits (0-9) representing spectral types and subtypes, LL luminosity classes, and PPPP represent spectral peculiarities. An entry "00" is returned for an unknown attribute. To take the example of the (incomplete) spectral classification just "A0", the LL codes will be "00" because the luminosity class is unspecified. The 4 spectral peculiarities are arranged into two subgroups of two each. Then, P1P2 are reserved for peculiairities that can be expected across the HR Diagram (e.g. "e," "p" or "n"), and P3P4 are codes for peculiarities common only to a smaller range of spectral types, e.g. types AF. In this way, one can represent as many as 4 spectral peculiarities, including the possible composite nature of the spectrum. The details of this have been published in an IVOA (International Virtual Observatory Alliance) Design Note, accessible at the url:

http://www.ivoa.net/Documents/latest/SpectClasses.html

## 9. Recent spectral classification catalogues

A very comprehensive catalogue of published spectral classifications has been developed by Brian A. Skiff of Lowell Observatory. Till 2008 it had 300,000 entries and was asymptotically complete up to about 1985 in the literature. It is a living catalogue in the sense that data is being included continuously.

The features of this catalogue are that the classifications in this compilation include only those types determined from spectra, omitting those determined from photometry. Classifications include MK types as well as types not strictly on the MK system, such as white dwarfs, Wolf–Rayet stars, etc. According to Brian A. Skiff, the catalogue includes for the first time results from many large-scale objective-prism spectral surveys done at Case, Stockholm, Crimea, Abastumani, and elsewhere. The stars in these surveys were usually identified only on charts or by other indirect means, and have been overlooked heretofore because of the difficulty in recovering the stars. Due to this reason, many of these stars are not included in SIMBAD. In addition primary MK standard stars are included from the lists of Morgan & Keenan, as well as from Garrison (1994) list of MK *anchor points.*

The catalogue description is given at URL: *http://cdsarc.u-strasbg.fr/viz-bin/Cat?B/mk*

Additional catalogues developed recently are listed below.

- *The Galactic O Star Catalog V.2.0* by Sota, A., Maíz Apellániz, J., Walborn, Nolan R., & Shida, R. Y. described in RMxAC, **33**, 56, 2008. It is a new version of O type star catalogue containing precise spectral classifications of O type stars brighter than 8th magnitude. The catalogue can be found at

http://www-int.stsci.edu/jmaiz/GOSmain.html

- *Contributions to the Nearby Stars (NStars) Project: Spectroscopy of Stars Earlier than M0 within 40 pc-The Southern Sample* Gray, R. O., Corbally, C. J., Garrison, R. F., McFadden, M. T., Bubar, E. J., McGahee, C. E., O'Donoghue, A. A., & Knox, E. R., 2006, AJ 132, 161

  This project is aimed at classifying the nearby solar-type stars (stars earlier than M0 within 40 pc of the sun). This work gives classification for 1676 stars south of the equator.

- *Visual Multiples. IX. MK Spectral Types*, Abt, H.A. 2008, ApJS 176, 216

  Classifications for 546 stars in multiple systems are given.

- *A Unified Near-Infrared Spectral Classification Scheme for T Dwarfs*, Burgasser, Adam J., Geballe, T. R., Leggett, S. K., Kirkpatrick, J. Davy, & Golimowski, David A., 2006, ApJ 637, 1067B

  This paper details a new unified near-infrared spectral classification scheme for T dwarfs, and presents spectral types for many of the known T dwarfs.

- Online spectral catalogs of T dwarfs are currently maintained by

  A. Burgasser (low resolution NIR: *http://www.browndwarfs.org/spexprism*;

  red optical: *http://web.mit.edu/∼ajb/www/tdwarf/#spectra*),

  S. Leggett (moderate resolution NIR:

  *http://www.jach.hawaii.edu/∼skl/LTdata.html*),

  I. McLean (moderate resolution NIR: *http://www.astro.ucla.edu/∼mclean/BDSSarchive/*),

  and J. Rayner (moderate resolultion NIR:

  *http://irtfweb.ifa.hawaii.edu/∼spex/WebLibrary/index.html#T*).

## 10.  Acknowledgment

It is a pleasure to thank Richard O. Gray for several electronic discussions and providing valuable information on recent classification catalogues. The Stellar Spectral classification Monograph by R.O. Gray and Chris Corbally has been extremely useful as it gives up-to-date information on the the spectral morphologies observed in different classes and the physical process related to them. I am also thankful to several investigators and publishers for allowing the usage of their materials (figures). I am grateful to Ranjan Gupta for pointing out several references .

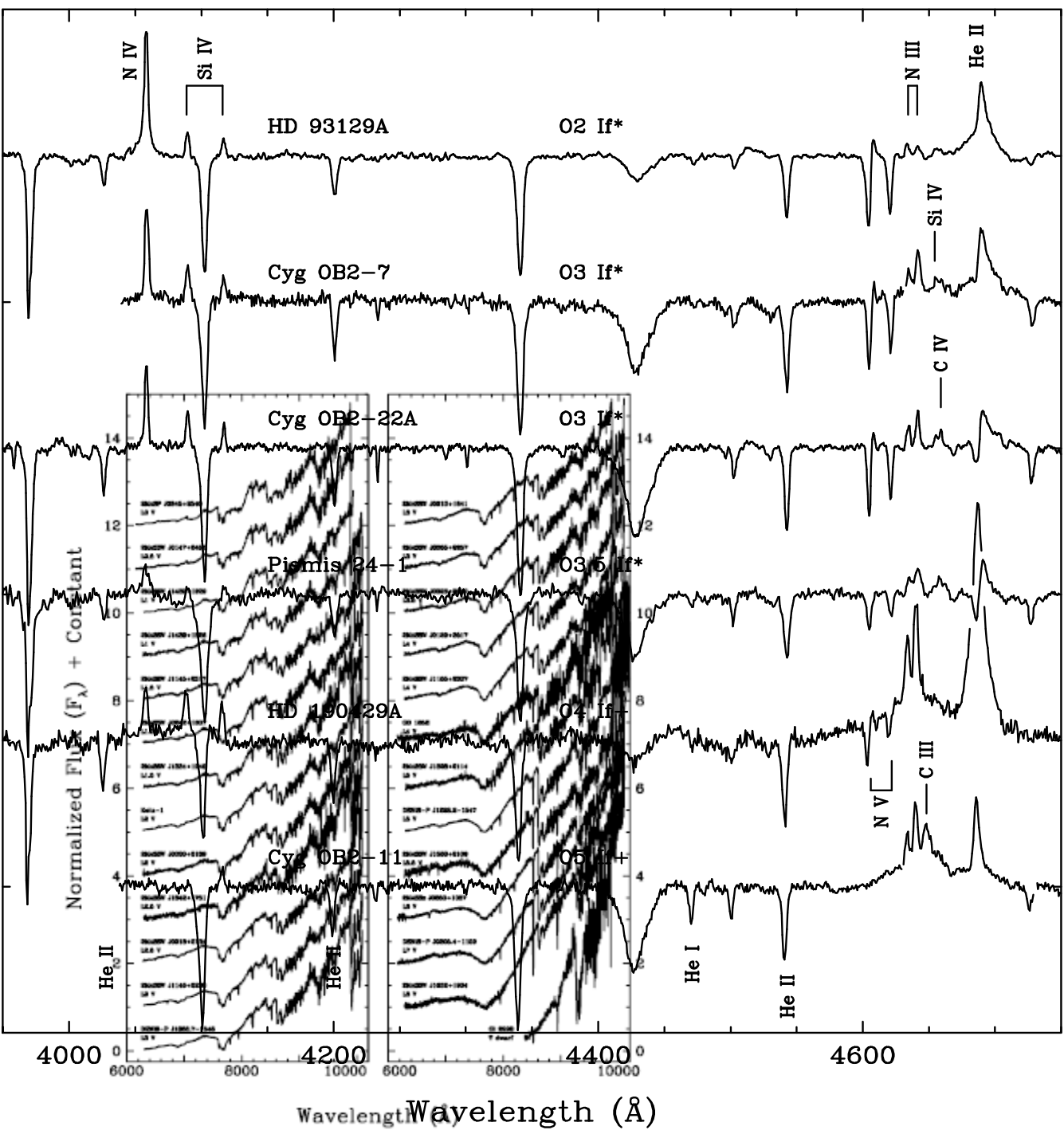

N IV
Si IV
HD 93129A    O2 If*
N III
He II

Cyg OB2-7    O3 If*
Si IV

Cyg OB2-22A    O3 If*
C IV

Pismis 24-1    O3.5 If*

HD 190429A    O4 If+

Cyg OB2-11    O5 If+
N V
C III

He I
He II

He II

He II    He II

Normalized Flux ($F_\lambda$) + Constant

Wavelength (Å)

4000    4200    4400    4600